\documentclass[twocolumn,amsmath,amssymb,pre]{revtex4-1}

\usepackage{graphicx}
\usepackage{color}
\usepackage{dcolumn}
\usepackage{amsmath}
\usepackage{CJK}
\usepackage{subfigure}
\usepackage{sidecap}
\usepackage{dsfont}

\usepackage{mathrsfs}
\usepackage{amsfonts}
\usepackage{indentfirst}
\usepackage{bm}

\usepackage{ulem}
\usepackage{xcolor}

\usepackage[colorlinks,citecolor=blue]{hyperref}

\newcommand{\be}{\begin{equation}}
\newcommand{\ee}{\end{equation}}
\newcommand{\bey}{\begin{eqnarray}}
\newcommand{\eey}{\end{eqnarray}}
\newcommand{\bw}{\begin{widetext}}
\newcommand{\ew}{\end{widetext}}

\newcommand{\ra}{\rangle}
\newcommand{\la}{\langle}

\newcommand{\ba}{\begin{array}}
\newcommand{\ea}{\end{array}}
\newcommand{\bi}{\begin{itemize}}
\newcommand{\ei}{\end{itemize}}
\newcommand{\bem}{\begin{enumerate}}
\newcommand{\eem}{\end{enumerate}}

\begin{document}

\title{Universal dynamics of the entropy of work distribution in spinor Bose-Einstein condensates}
\author{Zhen-Xia Niu}
\affiliation{Department of Physics, Zhejiang Normal University, Jinhua 321004, China}

\begin{abstract}

Driving a quantum many-body system across the quantum phase transition (QPT) in the finite time has been concerned in different branches of physics to explore various fundamental questions. 
Here, we analyze how the underlying QPT affects the work distribution $P(W)$, when the control parameter of a ferromagnetic spinor Bose-Einstein condensates is tuned through the critical point in the finite time.
We show that the work distribution undergoes a dramatic change with increasing the driving time $\tau$.
To capture the characteristics of the work distribution, we analyze the entropy of $P(W)$ and find three different regions in the evolution of entropy as a function of $\tau$.
Specifically, the entropy is insensitive to the driving time in the region of very short $\tau$, while it exhibits a universal power-law decay in the region with intermediate value of $\tau$. 
In particular, the power-law scaling of the entropy is  according with the well-known Kibble-Zurek mechanism.
For the region with large $\tau$, the validity of the adiabatic perturbation theory leads to the entropy decay as $\tau^{-2}\ln\tau$.   
Our results verify the usefulness of the entropy of the work distribution for understanding the critical dynamics and provide an alternative way to experimentally study nonequilibrium properties in quantum many-body systems.

\end{abstract}

\date{\today}

\maketitle

\section{Introduction}

The nonequilibrium dynamics in quantum many-body systems has attracted a lot of attention in statistical and condensed matter physics \cite{Polkovnikov2011,Eisert2015}.
The experimental progress has enabled the manipulation of nonequilibium dynamics in the cold atoms \cite{schreiber2015,Bordia2017} and ion traps \cite{zhang2017,de2021materials}.
The understandings of nonequilibrium systems are largely developed through analogies with  their thermal equilibrium counterparts.
Phase transitions, initially studied in equilibrium systems, signify the qualitative changes in properties of physical systems drived by tuning the control parameters. 
Notably, quantum systems at zero temperature may exhibit characteristics of a phase transition at some critical points of Hamiltonian control parameter, leading to the quantum phase transition (QPT) \cite{Mivehvar2024}. 
An intriguing feature of systems approaching a critical point is the breakdown of adiabaticity stemming from the vanishing energy gap, highlighting the intricate dynamics near critical points.
In this regard, the dynamical scaling near the quantum critical point 
\cite{Zurek2005,Dziarmaga2005,Dziarmaga2010,Campo2014,Acevedo2014,Defenu2018}, 
and the connection between the dynamical and equilibrium critical properties \cite{Polkovnikov2011} are two interesting topics. 
For the nonequilibrium dynamics driven by the sudden quench, the dynamical features resemble the behaviors of the thermodynamics functions at the critical point, 
which have been identified in both theoretical and experimental studies \cite{Zvyagin2016,Heyl2018}. 
However, how to link the dynamics of a QPT to the equilibrium critical phenomena for the slow quench is still elusive.

Driving an isolated quantum many-body system out of equilibrium is associated with the injection or extraction of the work during the nonequilibrium dynamical process.
In quantum mechanics, the work is a stochastic variable with significant fluctuations \cite{Talkner2007,Campisi2011}.
Consequently, the work for quantum systems is characterized by the work distribution, which encodes various information about the nonequilibrium dynamics \cite{Marino2014,Hoang2016,Campisi2017,Chenu2018,Goold2018,Abeling2016}.
Accordingly, the work distribution plays an important role in exploring the quantum critical dynamics \cite{Abeling2016,Silva2008,Campbell2016,QianW2017,Zawadzki2020,Zawadzki2023,Kiely2023}.
Recently, the signatures of the Kibble-Zurek  (KZ) mechanism  have been demonstrated in the work statistics, when a control parameter is driven across a QPT in the finite time \cite{Campo2014,Kibble1976,Zurek1985,Zurek1996,Damski2005,Puebla2020,FeiZ2020,FeiZ2021,ZhangF2022}. 
However, pervious works are mainly focused on the short-range interacting systems, such as various spin models, the situation for the long-range interacting systems remains less known. As the long-range interactions can strongly affect the critical and dynamical properties of quantum systems \cite{RevModPhys.95.035002}, it is therefore natural to ask what is the statistical properties of the work in quantum many-body systems with long-range interaction.

The work statistics in the system with infinite-range interaction has been explored in serveral very recent works \cite{gherardini2023,Andrea2024}. 
In the present work, we go a step further to investigate the critical features of the work statistics in a ferromagnetic spin-1 Bose-Einstein condensates (BEC) through the entropy of the work distribution. 
As a measure of the complexity of the work statistics, the entropy of the work distribution is useful for studying different phase transitions \cite{ZhangH2023}. 
By linearly tuning the control parameter of the system, we first discuss how the work distribution depends on the driving time. 
Then, a detailed analysis of the time dependence of entropy shows that the entropy undergoes a dramatic change with increasing driving time. 
In particular, in the slow quench, we demonstrate that the entropy of the work distribution exhibits a universal scaling, which is campatible with the KZ mechanism.

 \begin{figure*}
  \includegraphics[width=\textwidth]{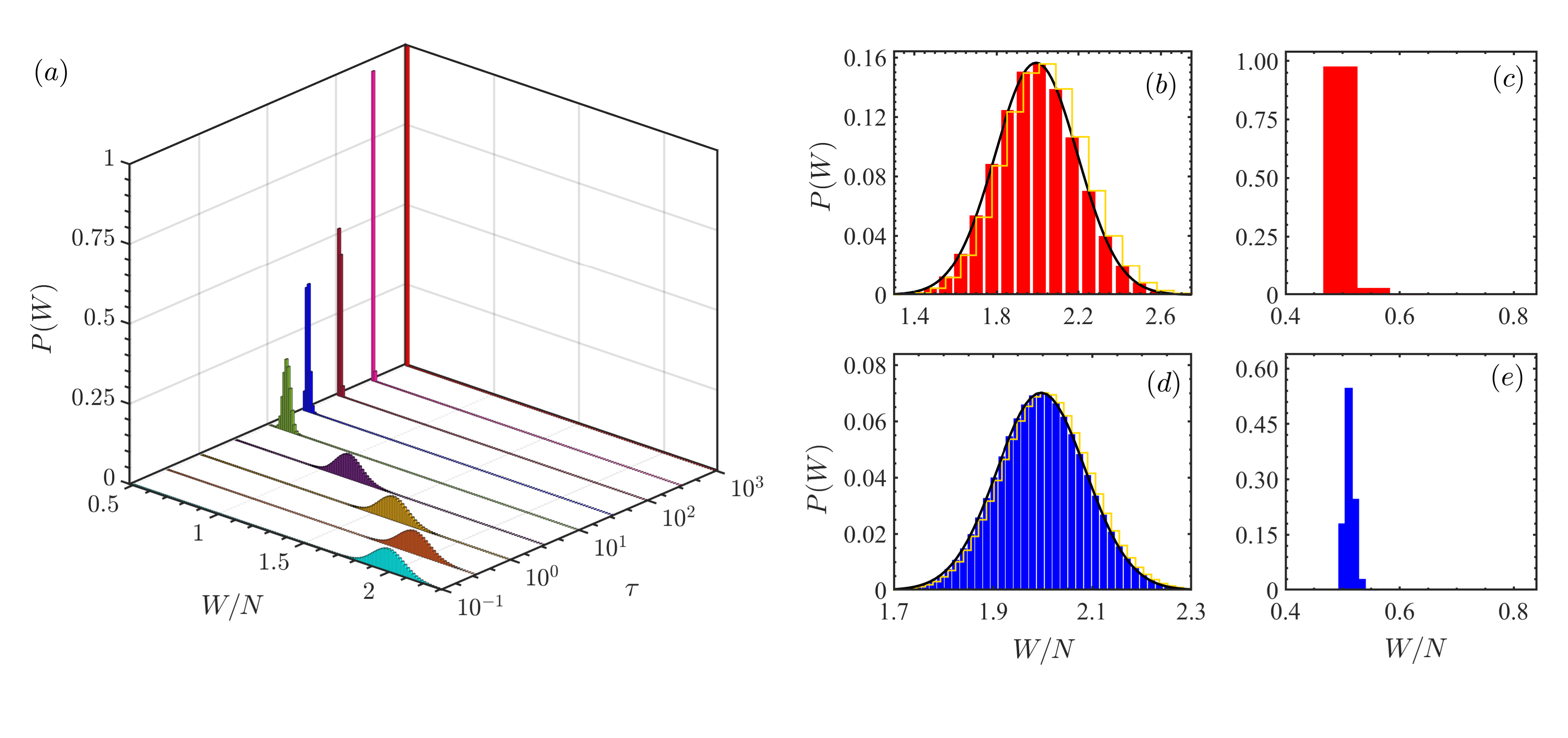}
  \caption{(a) Work distributions $P(W)$ in (\ref{SWPDF}) of the ferromagnetic spin-$1$ BEC for different driving times $\tau$ with $N=500$.
  (b)-(c): Full work distributions $P(W)$ with $N=100$ for $\tau=0.05$ (b) and $\tau=50$ (c).
  (d)-(e): Work distributions $P(W)$ for the same values of $\tau$ as in panels (b) and (c) with $N=500$.
  In panels (b) and (d), the yellow stairs show the work distributions for the sudden quench process, while the black solid curves denote the Gaussian distributions. 
  Other parameter: $\lambda_0=0$. All quantities are dimensionless.}
  \label{PdfWt}
 \end{figure*}

\section{Spin-$1$ Bose-Einstein condensates}

As a highly controllable platform, the spinor BEC \cite{Kawaguchi2012,Kurn2013} have been employed as a prototypical model, both in experiments and theoreties, for studying various quantum many-body phenomena, such as different phase transitions \cite{Sadler2006,Bookjans2011,TianT2020,Feldmann2021,ZhouL2023,Niu2023},
nonequilibrium dynamics \cite{ZhaoL2014,Anquez2016,XueM2018,Dag2018,ChenZ2019,LYQiu2020,Mittal2020,ZXNiu2023}, 
and quantum chaos \cite{ChengJ2010,Rautenberg2020,ChengJL2022,Evrard2023}, among others.
Here, we consider a BEC composed of $N$ spin-$1$ atoms, where the spin freedom decouples from the spatial mode. 
Using the single-mode approximation, the Hamiltonian of the system can be written as \cite{Feldmann2021,XueM2018,Dag2018}
\begin{align} \label{SpinH}
 \frac{H}{|c|}=&\frac{\mathrm{sign}(c)}{N}\left[(a_0^{\dag 2}a_1a_{-1}+a_{-1}^\dag a_1^\dag a_0^2)+N_0(N_{-1}+N_1)\right]\notag \\
 &+\lambda(N_{-1}+N_1),
\end{align}
where $a_m$ ($a_m^\dag$) is the bosonic annihilation (creation) operator for state $m=0,\pm1$.
$N_m=a_m^\dag a_m$ is the number operator of $m$th state and $\sum_mN_m=N$ is conserved.
$c$ is the strength of the spin-dependent interaction with $c<0\ (c>0)$,  corresponding to ferromagnetic (antiferromagnetic) interaction for $^{87}\mathrm{Rb}$ and $^{23}\mathrm{Na}$ atoms. 
$\lambda\equiv q/|c|$ represents the rescaled quadratic Zeeman shift, where the quadratic Zeeman shift $q$ can be tuned through the microwave dressing \cite{Gerbier2006,Hamley2012}. 
Moreover, the conservations of the total magnetization $M=N_1-N_{-1}$ and the parity $\Pi=(-1)^{N_0}$ further allow us to restrict our study in the subspace with $M=0$ and $\Pi=1$.  
Hence, the dimension of the Hilbert space is  $\mathcal{D}_\mathcal{H}=N/2+1$ for even $N$.

It is worth mentioning that the spin-$1$ BEC model is an all-to-all infinite-range interacting system, which provides an insights into the nonequilibrium exploration for the long-range interacting systems.
Therefore, studying the work distribution during a nonequilibrium process in spin-1 BEC could promote our understanding of the nonequilibrium dynamics in long-range interacting systems, which is an active research field in recent years \cite{Puebla2020b,DEFENU20241,RevModPhys.95.035002}.

In this work, we focus on the ferromagnetic interaction and positive quadratic Zeeman shift, so that $c<0$ and $\lambda\geq0$.
In the ferromagnetic spin-$1$ BEC described by the Hamiltonian (\ref{SpinH}), the system undergoes a QPT as $\lambda$ passes through the critical point $\lambda_c=2$, which separates the broken-axisymmetry phase with $0\leq\lambda<2$ from the polar phase with $\lambda>2$ 
\cite{Kawaguchi2012,Sadler2006,XueM2018,Feldmann2018}. 
The nonequilibrium dynamical properties of the system near the quantum critical point have been investigated in both experiment \cite{Anquez2016} and theoretics \cite{XueM2018,Saito2013}.
In particular, the KZ scaling \cite{Kibble1976,Zurek1985,Zurek1996} has been observed in the critical dynamics. 
To further reveal the impacts of the QPT on the nonequilibrium properties of the system, we explore how the quantum work statistics varies when driving the system across its critical point.

\section{Work distribution and its entropy}
The  work distribution $P(W)$ is a concept used in quantum mechanics to describe the work done on the system during a nonequilibrium process \cite{Talkner2007}.
For the isolated systems, $P(W)$ is usually obtained by two projective energy measurement on the initial and final Hamiltonians \cite{Talkner2007, Campisi2011}.
Specifically, we can assume that an isolated system is initially prepared in a state $\rho_0$ with Hamiltonian $H_0=\sum_nE_n^0|n\ra\la n|$.
Hence, the first energy measurement on $H_0$ gives the $n$th eigenstate $|n\ra$ with the energy $E_n^0$ and the probability $p_n^0=\la n|\rho_0|n\ra$.
At time $t=0$, the system starts to evolve according to a time-dependent Hamiltonian $H_t=\sum_kE_k^t|k_t\ra\la k_t|$ until $t=\tau$. 
Thus, the final state of the system is given by $\rho_\tau=U_{0\to\tau}\rho_0U^\dag_{0\to\tau}$, where $U_{0\to\tau}=\mathcal{T}\exp(-i\int_0^\tau H_t dt/\hbar)$ with the time order operator $\mathcal{T}$.
The outcome of the second energy measurement at $t=\tau$  is the $k$th eigenstate $|k_\tau\ra$ of 
$H_\tau=\sum_k E^\tau_k|k_\tau\ra\la k_\tau|$ with the probability $p^\tau_{k|n}=|\la k_\tau|U_{0\to\tau}|n\ra|^2$ and the energy $E_k^\tau$. 
Then, the work distribution during this evoluting process is given by
\be \label{WorkPDF}
  P(W)=\sum_{n,k}p_n^0p_{k|n}^\tau\delta[W-(E_k^\tau-E_n^0)].
\ee
It is worth pointing out that the probability $p_{k|n}^\tau$ becomes $p_{k|n}^\tau=|\la k_\tau|n\ra|^2$  for the sudden quench, while it reduces to $p_{k|n}^\tau=\delta_{k,n}$ in the adiabatic limit. 
Moreover, $P(W)$ can been experimentally extracted in different quantum systems \cite{Dorner2013,Mazzola2013,Roncaglia2014,Batalhao2014,Chiara2015}.

 \begin{figure}
  \includegraphics[width=\columnwidth]{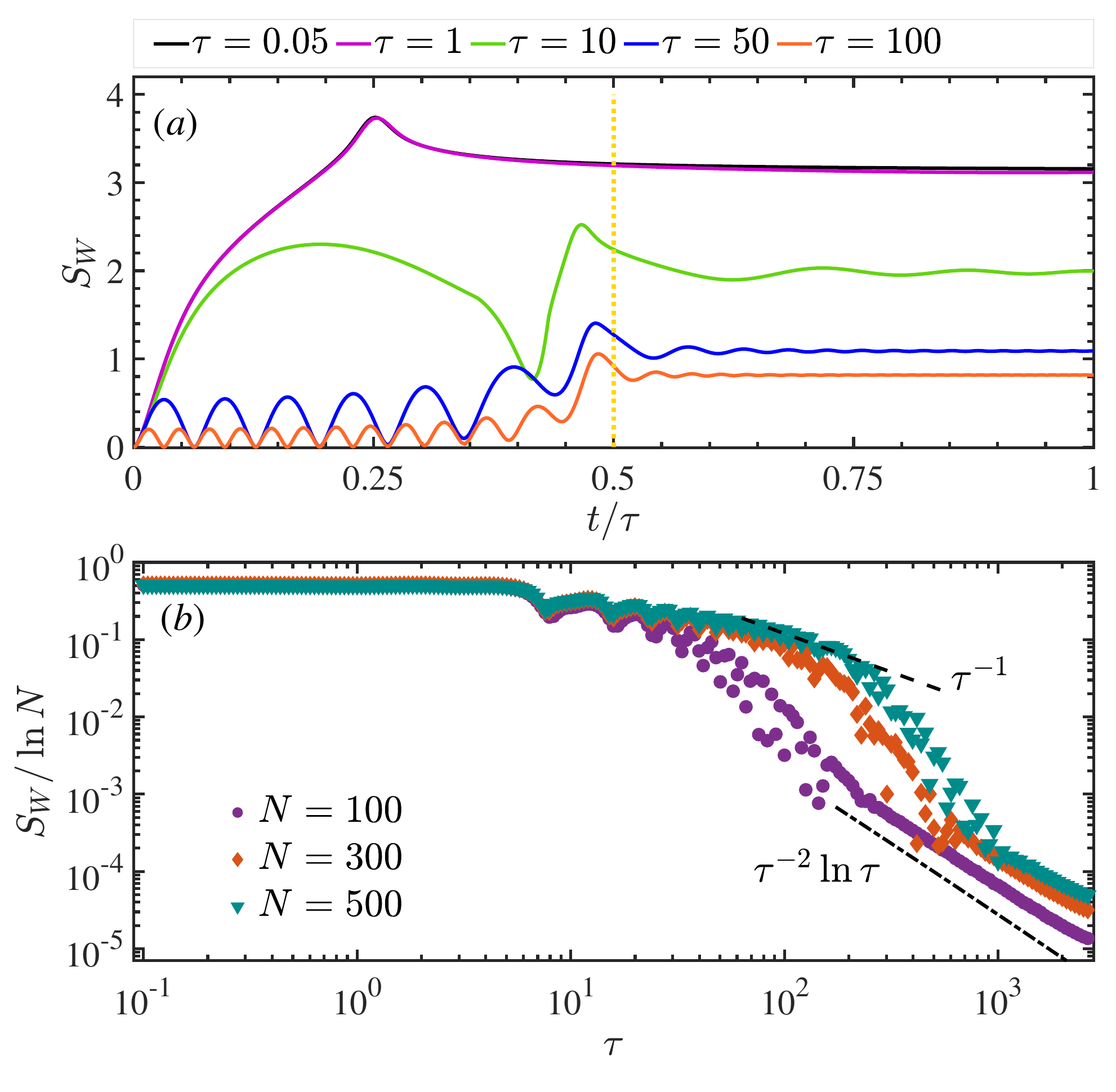}
  \caption{(a) Entropy $S_{W}$ of the work distribution $P(W)$  as a function of  the rescaled time $t/\tau$ for several driving times $\tau$ with system size $N=500$.
  The vertical yellow dotted line denotes the position of the critical point $\lambda_c$. 
  (b) Rescaled entropy, $S_W/\ln N$, as a function of driving time $\tau$ for different system sizes $N$. Here, the maximal value of the entropy is given by $S_{W,max}=\ln\mathcal{D}_\mathcal{H}$.
  The black dashed line denotes the power law $\tau^{-1}$, while the black dot-dashed line represents $\tau^{-2}\ln\tau$. Other parameter: $\lambda_0=0$. All quantities are dimensionless.}
  \label{SwTime}
 \end{figure}

In this spin-$1$ BEC model, we choose $\lambda$ as the control parameter and tune it across the critical point $\lambda_c=2$ by a linear drive
\be
 \lambda(t)=\lambda_0+2(\lambda_c-\lambda_0)t/\tau,
\ee
where $\tau$ is the driving time, with $\tau=0, \infty$ corresponding to the sudden and adiabatic quench, respectively.
And we consider that the system is initially prepared in the ground state of the Hamiltonian (\ref{SpinH}) with $\lambda_0=0$, 
so that $\rho_0=|{\mathrm{GS}_0}\ra\la\mathrm{GS}_0|$.
To this end, the work distribution in Eq.~(\ref{WorkPDF}) can be recast as
\be \label{SWPDF}
 P(W)=\sum_k p^\tau_{k|\mathrm{GS}}\delta[W-(E_k^\tau-E_\mathrm{GS}^0)],
\ee
where $p^\tau_{k|\mathrm{GS}}=|\la k_\tau|U_{0\to\tau}|\mathrm{GS}_0\ra|^2$.
The simplified work distribution, Eq.~(\ref{SWPDF}), is also known as the local density of states, which describes the energy distribution of the initial state over the final Hamiltonian spectrum.

It is necessary to know the evolving state $|\psi(t)\ra$ of system in nonequilibrium dynamics to obtain the full work distribution $P(W)$ in Eq.~(\ref{SWPDF}).
Thus, we expand $|\psi(t)\ra$ as $|\psi(t)\ra=\sum_k d_k(t)|k_t\ra$, where $|k_t\ra$ is the $k$th instantaneous eigenstates of $H[\lambda(t)]$ with eigenvalue $E_k^t$, so that $H[\lambda(t)]|k_t\ra=E_k^t|k_t\ra$. 
Then, inserting $|\psi(t)\ra$ into the time-dependent Schr\"{o}dinger equation $i\partial_t|\psi(t)\ra=H[\lambda(t)]|\psi(t)\ra$, it is straightforward to find that
\be\label{Evexpcf}
 i\partial_td_k(t)+i\sum_{k^{\prime}}d_{k^{\prime}}(t)\la k_t|\partial_tk^{\prime}_t\ra=E_k^td_k(t).
\ee
By performing a gauge transformation ${d}_k(t)=e^{-i\beta_k(t)}{\alpha}_k(t)$ with the dynamical phase $\beta_k(t)=\int_0^t E_k(s)ds$, one can rewrite the Eq.~(\ref{Evexpcf}) as follows
\be\label{Evexpfc}
  \partial_t\alpha_k(t)=-\sum_{k^{\prime}}e^{i[\beta_k(t)-\beta_{k^{\prime}}(t)]}\alpha_{k^{\prime}}(t)
        \la k_t|\partial_{t}k^{\prime}_{t}\ra.
\ee
We numerically solve the Eq.~(\ref{Evexpfc}) by the exact diagonalization with the initial condition $\alpha_k(0)=\delta_{k1}$.
The state at $t=\tau$ is given by $|\psi(\tau)\ra=\sum_k d_k(\tau)|k_\tau\ra$.
Then the transition probability is  $p_{k|\mathrm{GS}}^\tau=|\la k_\tau|\psi(\tau)\ra|^2=|d_k(\tau)|^2=|\alpha_k(\tau)|^2$.

In Fig.~\ref{PdfWt}(a), we plot the full work distributions, Eq.~(\ref{SWPDF}), of the ferromagnetic spin-$1$ BEC for several driving time $\tau$ and system sizes $N$.
We find that the work distribution $P(W)$ undergoes a dramatic change with increasing the driving time $\tau$, regardless of the system size.
For $\tau\lesssim1$, $P(W)$ exhibits an obvious resemblance to the Gaussian distribution with a peak located around the average $\langle W\rangle=\int WP(W)dW$, as clearly shown in Figs.~\ref{PdfWt}(b) and \ref{PdfWt}(d).
Moreover, the work distribution $P(W)$ is independent of the driving time $\tau$ for such fast ramping drive.
This implies that fast drive, but not instantaneous quench, is still within the validity of sudden quench approximation.
As the driving time $\tau$ is increased, $P(W)$ shifts to the small values of the work and becomes increasingly asymmetric.
For the slow drive, such as the cases of $\tau=50$ plotted in Figs.~\ref{PdfWt}(c) and \ref{PdfWt}(e),
the work distributions $P(W)$ evolve into the single peak, which correspondes to the adiabatic ground state. 

With increasing the system size $N$, excitation gap is decreased \cite{XueM2018}, which leads to a large amount of excitations emerging in the final state.
As a consequence, the Gaussianity of $P(W)$ is enhanced with increasing the system size $N$, as demonstrated in Fig.~\ref{PdfWt}(d).
However, the decrease of the excitation gap increases the driving time, when the system adiabatically passes through the pseudocritical point of the finite-size system.

To quantitatively characterize the signatures of $P(W)$, a common way is to study the summary statistics, such as moments or cummulants of $P(W)$.
Here, we focus on the entropy $S_W$ of $P(W)$, $S_W=-\sum_WP(W)\ln P(W)$, which was first introduced and defined in Ref.~\cite{Kiely2023} to 
measure the complexity of the work distribution. 
The variation of  $S_W$ is in the interval $S_W\in[0,\ln\mathcal{D}^2_\mathcal{H}]$. 
Here two extreme values correspond to the deterministic and uniform work distributions, respectively.

For the sudden quench in quantum many-body systems, the usefulness of $S_W$ for understanding the different phase transitions has been verified in recent studies \cite{Kiely2023, ZhangH2023}.
However, as the control parameter through the critical point of QPTs in the finite time, the dynamics and scaling properties of $S_W$ have not yet been explored.

Duing to the energy levels of this system in the case with even parity are nondegenerate, the entropy of $P(W)$ can be simplified as
\be \label{SSW}
  S_W=-\sum_k p_{k|\mathrm{GS}}^\tau\ln p_{k|\mathrm{GS}}^\tau.
\ee 
Here, we take the ground state of the system as the initial state, so that the maximal value of $S_W$ in Eq.~(\ref{SSW}) is $\ln\mathcal{D}_\mathcal{H}$.
According to the indicated features of the work distribution $P(W)$ in Fig.~\ref{PdfWt}, the entropy $S_W$ may undergo a dramatic change for different quench rates $\dot{\lambda}(t)$, in which the signatures of the critical dynamics can be characterized by $S_W$. 

In Fig.~\ref{SwTime}(a), we plot $S_W$ as a function of the rescaled time $t/\tau$ for several driving times $\tau$ with $N=500$.
For the very short driving times with $\tau\ll 1$, the sudden quench approximation implies that the system is frozen at the initial state during the whole quench process.
As a consequence, the entropy of the work distribution shows a rapid growth with time, and then decreases to its saturation value at long time, regardless of the values of $\tau$. 
The presence of the QPT is unveiled by the peak of $S_W$, which can be regarded as the finite-size precursor of the QPT. 
With increasing the driving time $\tau$, the evolution of $S_W$ shows different oscillation patterns.
Specifically, for larger $\tau$, such as the case of $\tau=100$, $S_W$ presents a high-frequency oscillation away from the critical point.
While the oscillations slow down near the critical point. 
In addition, the oscillation has a large envelope around the critical point, comparing to the evolutions away from the critical point.

Different oscillation behaviors in $S_W$ for large $\tau$ can be understood by the adiabatic perturbation theory, which gives the excitation probability as \cite{Grandi2010}
\begin{align}\label{TrP}
 p_{k|\mathrm{GS}}\approx&|\alpha(\lambda_\tau)|^2 \notag \\
 =&\tau^{-2}\left\{\frac{|\la k_\tau|\partial_{\lambda_0}|\mathrm{GS}_0\ra|^2}
 {(E_k^0-E_{\mathrm{GS}}^0)^2}+\frac{|\la k_\tau|\partial_{\lambda_{\tau}}|\mathrm{GS}_0\ra|^2}
 {(E_k^\tau-E_{\mathrm{GS}}^\tau)^2} \right. \notag \\
 &\left.-\frac{2\la k_\tau|\partial_{\lambda_0}|\mathrm{GS}_0\ra\la k_\tau|\partial_{\lambda_\tau}|\mathrm{GS}_0\ra
 \cos\left(\Delta\beta_{kg}\right)}
 {(E_k^0-E_{\mathrm{GS}}^0)(E_k^\tau-E_{\mathrm{GS}}^\tau)}
 \right\}.
\end{align}
Here, $\lambda_\tau = \lambda(\tau)$, 
$\Delta\beta_{kg}=\beta_k(\lambda_\tau)-\beta_g(\lambda_\tau)-\beta_k(\lambda_0)+\beta_g(\lambda_0)$ denotes the accumulated phase difference between 
the $k$th excited state and the ground state during the driving process. 
In the adiabatic limit, as only the transition to the first excited state needs to be considered, 
we thus have $\Delta\beta_{kg}=\Delta\beta_{21}\simeq\int_{\lambda_0}^{\lambda_\tau}\Delta E(s)ds$,
with $\Delta E(s)=E_2(s)-E_1(s)$ being the energy gap between the first excited and ground states. 
The energy gap usually has large values away from the critical point, which gives rise to a high-frequency oscillation  in the excitation probability.
Thus, the entropy $S_W$ undergoes fast oscillations.
On the contrary, the gap narrows near the critical point, which leads to a slow oscillation in the evolution of $S_W$.  

To further reveal the dynamical properties of $S_W$, we capture how the rescaled entropy $S_W/\ln N$ varies as a function of the driving time $\tau$ for several system sizes $N$ in Fig.~\ref{SwTime}(b).
One can clearly see that the dependence of $S_W$ on $\tau$ shows three distinct regions.
The first region correspondes to the short driving time ($\tau\lesssim10$),  and we call it fast quench region.
In this region, due to the fact that the short driving time leads to the freeze of the system state during the time evolution, the rescaled entropy remains unchange with increasing $\tau$. 
Moreover, we also observe that the value of the rescaled entropy is almost independent of the system size.
It is worth pointing out that the dynamics is nonuniversal in the fast quench region.

Following the fast quench region, the entirely different behaviors are observed in the intermediate region, where the rescaled entropy decreases with increasing $\tau$.
The numerical fitting reveals that the decay of the rescaled entropy can be captured by a power-law form $\tau^{-1}$, as indicated by the black dashed line in Fig.~\ref{SwTime}(b).
The third region correspondes to the large $\tau$, named as the adiabatic region, in which the adiabatic perturbation limit is valid.
The rescaled entropy also decreases as increasing $\tau$ in adiabatic region.
While its decay behavior  can be well described by $\tau^{-2}\ln\tau$ scaling [see the black dot-dashed line in Fig.~\ref{SwTime}(b)], and distinct from that in the intermediate region.

For the finite-size system, even though the energy gap remains finite, the system can undergo a nonadiabatic evolution when it is driven across the critical point. 
To observe the diabatic effects, the driving rate should be fast such that the ground state get excited.  
This is ensured for our considered intermediate region. In this region, the system dynamics is governed by the so-called  KZ time 
$\hat{t}_{KZ}\sim\tau^{z\nu/(1+z\nu)}$ or KZ length scale $\hat{\xi}_{KZ}\sim\tau^{\nu/(1+z\nu)}$\cite{Zurek1996,XueM2018}, around the critical point.
Here, $\nu$ and $z$ are the spatial and dynamical critical exponents, respectively.
Then, one can obtain the number of excitation $n_{ex}=1/\hat{\xi}_{KZ}^d\sim\tau^{-d\nu/(1+z\nu)}$ 
with $d$ being the upper critical dimensionality of the system.
One can expect that a characteristic amount of entropy $s_w$ is associated with each excitation, so that $S_W\sim n_{ex}s_w\sim\tau^{-d\nu/(1+z\nu)}$.
This is the KZ scaling of $S_W$ and hold in the limit of $\tau\to\infty$ in the thermodynamic limit $N\to\infty$. 
For our considered system, it is known that $\nu=1/2$, $z=1$, and $d=3$, we therefore have $S_W\sim\tau^{-1}$, as plotted by the black dashed line in Fig.~\ref{SwTime}(b). 

The finite energy gap of a finite-size system also implies that the control parameter can be adiabatically tuned across the critical point for large enough
driving time $\tau\gg1$.  
This leads to the presence of the adiabatic region and the entropy $S_W$ in this region can be approximated as
\be
  S_W\simeq p_{1|\mathrm{GS}}^\tau\ln p^\tau_{1|\mathrm{GS}}.
\ee
Here, $p_{1|\mathrm{GS}}^\tau$ is the probability of the transition between the initial state and the ground state of the final Hamitonian.  
According to the Eq.~(\ref{TrP}), we know that $p_{1|\mathrm{GS}}^\tau\sim\tau^{-2}$ and thus $S_W\sim\tau^{-2}\ln\tau$. 
As illustrated in Fig.~\ref{SwTime}(b), the numerical results in the large-$\tau$ (adiabatic) region are good agreement with this scaling behavior of $S_W$.

\section{conclusions}

Driving a quantum many-body system across its critical point in the finite time is usually accompanied by the energy excitations.
In order to the underlying mechanism of the nonequilibrium critical phenomenon can shed light on several fundamental questions that arise in both theoretical and experimental studies.
In this work, we address the question of how the QPT affects the nonequilibrium dynamics by the work statistics in the ferromagnetic spin-$1$ BEC driven in the finite time.
Although the critical properties of the work statistics for various spin models with short-range interaction have been examined in several works \cite{Zawadzki2020,Zawadzki2023,Kiely2023,FeiZ2020,FeiZ2021,ZhangF2022}, the situation for long-range interacting quantum systems is still less known.
By taking the quadratic Zeeman shift as the control parameter, we have explored the dynamics from sudden quench regime to adiabatic regime by linearly tuning the control parameter across the critical point.
We have found that the work distribution $P(W)$ exhibits a strong dependence on the driving rate. 
And the Gaussianity of $P(W)$ is decreased with increasing the driving time.
To quantitatively capture the characteristics of $P(W)$, we have examined how the entropy $S_W$ of the work distribution varies as a function of the driving time $\tau$.
Three distinct dynamical regions, named as fast quench, intermediate, and adiabatic regions, are found in the evolution of $S_W$.
The behavior of $S_W$ in the fast quench region is nonuniversal and independent of $\tau$, while it shows universal scaling behavior in other two dynamical regions.
We have demonstrated that the universal scaling of $S_W$ in the intermediate and adiabatic regions can be explained by the Kibble-Zurek mechanism and adiabatic perturbation theory, respectively.

Our findings help us to get further understanding for the signatures of  critical dynamics in nonequilibrium systems.
A natural extension of the present work is to consider the thermal initial state and discuss the effects of finite temperture in the critical dynamics.
Another interesting topic for future exploration is to analyze the scaling behavior of $S_W$ in the systems with short-range interaction.
Moreover, how the nonlinear quenches affect the scaling behavior of the work statistics is also deserved for future exploration.
It was known that the nonlinear ramps have nontrivial impacts on the nonequilibrium dynamics \cite{Puebla2020b,Diptiman2008,Barankov2008}.
One can therefore expect that the scaling behavior of the work distribution and its entropy should undergo a dramatic change for the nonlinear quench.  
Finally, as the spin-$1$ BEC is a highly controllable platform and the work distribution can be measured by current experimental technics, we expect that our studies could motivate more expermental investigation of the critical dynamics in quantum many-body systems. 

\acknowledgements

We would like to thank Qian Wang for insightful discussions.
This work was supported by the Zhejiang Provincial Nature Science Foundation under 
Grant No.~LQ22A040006.

\bibliographystyle{apsrev4-1}
%

\end{document}